%% file: main.tex
\documentclass[aps,prx,superscriptaddress,twocolumn,10pt]{revtex4-2}

\makeatletter
\renewcommand{\bibinfo}[2]{%
  \ifstrequal{#1}{title}{\textit{#2}}{#2}}
\makeatother
\usepackage[titletoc,toc,title,page]{appendix}

\usepackage{csquotes}
\usepackage[italian,english]{babel}

\usepackage{microtype} 
\usepackage[margin=.6in]{geometry}

\usepackage[sc]{mathpazo} 

\usepackage{bm} 

\usepackage{dsfont} 
\usepackage{amsmath,amssymb,amsthm,thmtools}
\usepackage{mathtools}
\usepackage{cases}
\usepackage{calc}
\usepackage{mathrsfs} 
\usepackage[normalem]{ulem} 
\usepackage{nameref}
\usepackage[colorlinks=true]{hyperref}
\usepackage[nameinlink]{cleveref}
\crefname{appsec}{Appendix}{Appendices}
\crefname{box}{Box}{Box}
\hypersetup{
  colorlinks   = true, 
  urlcolor     = green!80!black, 
  linkcolor    = blue, 
  citecolor    = red!80!black, 
  breaklinks=true   
}

\usepackage{physics} 

\usepackage{float} 

\usepackage{graphicx}

\usepackage[usenames,dvipsnames,table]{xcolor}
\usepackage{easyReview}

\usepackage{tikz}
\usetikzlibrary{calc,shapes.geometric}

\usepackage{multirow,tabularx,booktabs}
\usepackage{makecell}

\newcommand{\calO}{\mathcal{O}}

\newcommand{\bs}[1]{\boldsymbol{#1}}
\newcommand{\on}[1]{\operatorname{#1}}
\newcommand{\parTitle}[1]{\noindent\emph{#1} --- }

\newcommand{\bstildemu}{\bs{\tilde{\mu}}}

\definecolor{azure}{rgb}{0.0, 0.3, 1.0}

\makeatletter
\newcommand{\labeltarget}[1]{\Hy@raisedlink{\hypertarget{#1}{}}}
\makeatother
\begin{document}

\title{Experimental property-reconstruction in a photonic quantum extreme learning machine}
\author{Alessia Suprano}\thanks{These authors contributed equally to this work}
\affiliation{Dipartimento di Fisica - Sapienza Università di Roma\comma{} P.le Aldo Moro 5\comma{} I-00185 Roma\comma{} Italy}
\author{Danilo Zia}\thanks{These authors contributed equally to this work}
\affiliation{Dipartimento di Fisica - Sapienza Università di Roma\comma{} P.le Aldo Moro 5\comma{} I-00185 Roma\comma{} Italy}
\author{Luca Innocenti}\thanks{These authors contributed equally to this work}
\affiliation{Universit\`a degli Studi di Palermo\comma{} Dipartimento di Fisica e Chimica - Emilio Segr\`e\comma{} via Archirafi 36\comma{} I-90123 Palermo\comma{} Italy}
\author{Salvatore Lorenzo}\thanks{These authors contributed equally to this work}
\affiliation{Universit\`a degli Studi di Palermo\comma{} Dipartimento di Fisica e Chimica - Emilio Segr\`e\comma{} via Archirafi 36\comma{} I-90123 Palermo\comma{} Italy}
\author{Valeria Cimini}
\affiliation{Dipartimento di Fisica - Sapienza Università di Roma\comma{} P.le Aldo Moro 5\comma{} I-00185 Roma\comma{} Italy}
\author{Taira Giordani}
\affiliation{Dipartimento di Fisica - Sapienza Università di Roma\comma{} P.le Aldo Moro 5\comma{} I-00185 Roma\comma{} Italy}
\let\comma,
\author{Ivan Palmisano}
\affiliation{Centre for Quantum Materials and Technologies\comma{} School of Mathematics and Physics\comma{} Queen's University Belfast\comma{} BT7 1NN\comma{} United Kingdom}
\author{Emanuele Polino}
\affiliation{Dipartimento di Fisica - Sapienza Università di Roma\comma{} P.le Aldo Moro 5\comma{} I-00185 Roma\comma{} Italy}
\affiliation{Centre for Quantum Dynamics and Centre for Quantum Computation and Communication Technology\comma{} Griffith University\comma{}  Yuggera Country \comma{}   Brisbane\comma{} Queensland 4111\comma{} Australia}
\author{Nicol\`{o} Spagnolo}
\affiliation{Dipartimento di Fisica - Sapienza Università di Roma\comma{} P.le Aldo Moro 5\comma{} I-00185 Roma\comma{} Italy}
\author{Fabio Sciarrino }
\affiliation{Dipartimento di Fisica - Sapienza Università di Roma\comma{} P.le Aldo Moro 5\comma{} I-00185 Roma\comma{} Italy}
\let\comma,
\author{G. Massimo Palma}
\affiliation{Universit\`a degli Studi di Palermo\comma{} Dipartimento di Fisica e Chimica - Emilio Segr\`e\comma{} via Archirafi 36\comma{} I-90123 Palermo\comma{} Italy}
\author{Alessandro Ferraro}
\affiliation{Centre for Quantum Materials and Technologies\comma{} School of Mathematics and Physics\comma{} Queen's University Belfast\comma{} BT7 1NN\comma{} United Kingdom}
\affiliation{Quantum Technology Lab\comma{} Dipartimento di Fisica Aldo Pontremoli\comma{} Universit\`a degli Studi di Milano\comma{} I-20133 Milano\comma{} Italy}
\author{Mauro Paternostro}
\affiliation{Centre for Quantum Materials and Technologies\comma{} School of Mathematics and Physics\comma{} Queen's University Belfast\comma{} BT7 1NN\comma{} United Kingdom}

\begin{abstract}
Recent developments have led to the possibility of embedding machine learning tools into experimental platforms to address key problems, including the characterization of the properties of quantum states.
Leveraging on this, we implement a quantum extreme learning machine in a photonic platform to achieve resource-efficient and accurate characterization of the polarization state of a photon. The underlying reservoir dynamics through which such input state evolves is implemented using the coined quantum walk of high-dimensional photonic orbital angular momentum, and performing projective measurements over a fixed basis. We demonstrate how the reconstruction of an unknown polarization state does not need a careful characterization of the measurement apparatus and is robust to experimental imperfections, thus representing a promising route for resource-economic state characterisation.
\end{abstract}

\maketitle

\parTitle{Context \& Motivations}
Accurate and resource-efficient estimation of properties of quantum states is a pivotal task in quantum information science, particularly in areas such as quantum metrology~\cite{polino2020photonic,giovannetti2011advances,czerwinski2022selected,gebhart2023learning}.
In particular, estimation strategies relying on single measurement settings have attracted notable attention in the last years~\cite{stricker2022experimental,bian2015realization,galvis2023single,struchalin2021experimental}.
Significant attention has also been devoted to the theoretical analysis of state estimation protocols based on randomized measurements, in particular through shadow tomography protocols~\cite{huang2020predicting,zhou2022performance,elben2023randomized,tran2023measuring,stricker2022experimental}, which were later shown to be applicable in generic measurement scenarios~\cite{acharya2021informationally,innocenti2022potential,nguyen2022optimising}.
On the other hand, several works have demonstrated the usefulness of integrating machine learning tools to implement and enhance the efficiency of quantum state estimation strategies~\cite{giordaniVVB,suprano2021dynamical,Suprano2021,zia2023,lohani2020machine,melnikov2023quantum, rocchetto2019, Santagatieaap9646,wangpaesani,carrasquilla2019reconstructing,cimini2023deep, palmieri2020experimental, ahmed2021quantum}.
In particular, Quantum Extreme Learning Machines (QELMs)~\cite{mujal2021opportunities,huang2011extreme} have been proposed as a particularly efficient medium to extract features from input quantum states with a flexible architecture~\cite{ghosh2019quantum,krisnanda2023tomographic,innocenti2022potential}.

In this work, we leverage QELMs to efficiently recover properties of photonic quantum states encoded in the polarization degree of freedom, exploiting orbital angular momentum (OAM) as an ancillary degree of freedom to enable reconstruction via a single measurement setting.
The interaction between polarization and OAM, which is experimentally implemented via a quantum-walk-based photonic apparatus~\cite{innocenti2017quantum,giordani2019experimental}, allows to extract information about the input polarization state by only measuring the OAM of the final state.
In the context of QELMs, the evolution mapping input polarization to output OAM takes the role of ``reservoir dynamics'', and enables complete reconstruction using a single measurement basis.
Using the framework of QELMs has the significant advantage of enabling the retrieval of information about the input state even without complete knowledge of the experimental apparatus itself.
This makes for an extremely flexible platform to extract features of input states, and is in stark contrast with conventional reconstruction pipelines, which crucially rely on accurate models of the evolution and measurement undergone by the states.
QELMs operate effectively without this assumption, requiring only access to a training dataset of known states —-- a task that is often less demanding in practice.
While experimental demonstrations of single-setting quantum state estimation have been reported in a few different platforms~\cite{stricker2022experimental,bian2015realization,galvis2023single}, reconstructions in all such protocols rely on accurate prior knowledge of all parts of the experimental apparatus.
By contrast, our QELM-based strategy makes for a highly flexible estimation strategy, resilient to many types of experimental noise and misalignment, by virtue of the training stage automatically adapting the post-processing to enable accurate reconstruction.
We benchmark our results with the accuracies obtained using alternative estimation strategies, finding our QELM-based approach to clearly outperform the alternatives in the considered scenarios.
It is worth noting that the reported reconstruction strategy is fully general and applicable to any experimental scenario where the goal is reconstructing properties of input states, even though only a partially characterized measurement stage is available. Furthermore, as discussed in depth in Refs.~\cite{innocenti2022potential,innocenti2023shadow}, the statistics required for accurate reconstruction mostly depend on the symmetry properties of the effective measurement implemented by the setup, rather than the dimension of the state one wishes to reconstruct.

\parTitle{QELM estimation framework}
QELMs operate by exploiting an uncharacterized time-independent dynamic to extract target properties from input states. To achieve this, the scheme uses a training dataset of quantum states to figure out the best way to extract the sought-after features from the measurement data~\cite{innocenti2022potential}.
The use of a training dataset allows one to forgo the need to characterize the measurement apparatus itself: the training process automatically adjusts to the complexities of the experimental reality.
Furthermore, training QELMs is a particularly simple endeavor, amounting to solving a linear regression problem, and is, therefore, less prone to overfitting issues, especially when used to extract linear features such as expectation values of observables~\cite{innocenti2022potential}.
More formally, a QELM involves evolving input states $\rho$ through some quantum channel $\Phi$ --- giving rise to what we will refer to as {\it reservoir dynamics} hereafter --- and then measuring them with some Positive Operator-Valued Measure (POVM) $\boldsymbol\mu\equiv (\mu_b)_b$.
Using a training dataset of the form $\{(\bm p_k^{\rm tr}, o_k)\}_k$ with $\bm p_k^{\rm tr}$ the probability vector resulting from measuring $\rho_k^{\rm tr}$, $(\bm p_k^{\rm tr})_b\equiv \trace(\mu_b \rho_k^{\rm tr})$, and $o_k\equiv \trace(\mathcal O\rho_k^{\rm tr})$ for some target observable $\mathcal O$, one can find a linear transformation $\bm w\equiv (w_b)_b$ such that $\sum_b w_b \trace(\mu_b \rho)\approx \trace(\mathcal O\rho)$ for all $\rho$.
In words, finding this $\bm w$ allows to read the target expectation values of input states directly from the measurement data, without requiring knowledge on the dynamic $\Phi$ and on the POVM $\boldsymbol\mu$ themselves (see Supplemental Material for more details).
This protocol can be seamlessly adapted to the case of multiple target observables.
The expressivity of a QELM --- that is, the space of observables that can be accurately retrieved for a given choice of $\Phi$ and $\boldsymbol\mu$ --- was proven to depend exclusively on the properties of the ``effective POVM'', that is, the POVM with elements $\tilde\mu_b\equiv \Phi^\dagger(\mu_b)$, where $\Phi^\dagger$ denotes the adjoint of $\Phi$~\cite{watrous2018theory}.
In particular, a necessary condition for enabling the reconstruction of arbitrary observables is that the reservoir dynamic $\Phi$ must enlarge the dimension of the input space in order to guarantee a sufficiently large number of measurement outcomes~\cite{innocenti2022potential}.

\begin{figure}[ht!]
    \centering
    \includegraphics[width=\columnwidth]{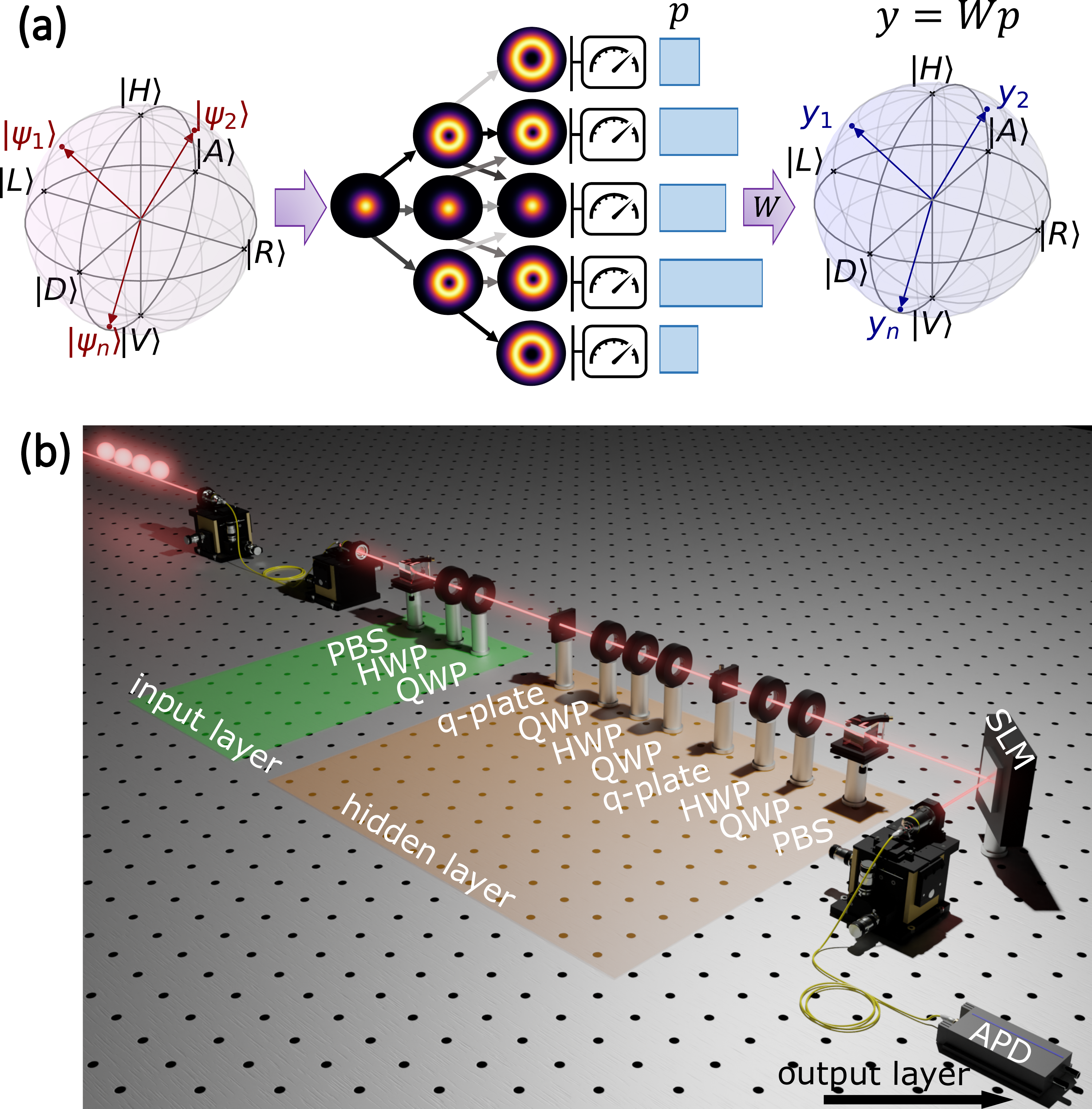}
    \caption{\textbf{Experimental QELM}. 
    \textbf{(a)} Schematic overview of the experimental QELM. Initial quantum states $|\psi_1 \rangle,|\psi_2 \rangle, \cdots, |\psi_n \rangle$ encoded in the polarization degree of freedom of single photons evolve through a reservoir dynamic, in which the information encoded in the initial two-dimensional space is transferred into the larger Hilbert space of the OAM. By performing only projective measurements on the OAM computational basis, the QELM is trained to reconstruct a set of target values $y_1,y_2, \cdots, y_n$. 
    \textbf{(b)} Experimental implementation. Single photons, generated at 808 nm via spontaneous parametric down-conversion, are sent through the state-preparation stage ({\it input layer}) made by a Polarizing-Beam Splitter (PBS), a Half-Wave Plate (HWP) and a Quarter-Wave Plate (QWP) to encode the initial state in the polarization degree of freedom. Subsequently, 
    the input states evolve through the {\it hidden layer} following the quantum walk dynamics implemented by HWPs, QWPs, and Q-Plates (QPs). After projecting onto the polarization state $|\psi_{\rm pol}\rangle$ with a sequence of HWP, QWP, and PBS, projective measurements in the OAM computational basis, $\mathcal{B}=\{\ket{n}\}$ with $n=\{-2,..,2\}$, are performed through a Spatial Light Modulator (SLM) followed by the coupling into a single-mode fiber. From the counts measured by an Avalanche Photodiode (APD), the output layer of the QELM is trained to retrieve the expectation values of the observables $\{\sigma_x, \sigma_y,\sigma_z \}$.}
    \label{fig:exp_setup}
\end{figure}

\parTitle{Experimental estimation strategy}
We implement experimentally the QELM-based quantum state estimation strategy using as reservoir dynamic a coined quantum walk (QW) in polarization and orbital angular momentum (OAM) of single photons~\cite{innocenti2017quantum,giordani2019experimental}.
The goal of the protocol is to extract expectation values of observables on the input polarization states, using the reservoir dynamics to transfer this information into the larger OAM space that is then measured (see~\cref{fig:exp_setup}\textbf{a}).
More specifically, we use states of the form
$ \vert{\Psi_f}\rangle =
\left(\prod_{k=1}^s S C_k\right)
\ket{0,\psi}$,
with $C_k\equiv I\otimes U_k$ the unitary \textit{coin operation}, acting nontrivially only on the coin space, and with
$S \equiv \sqrt{p} (I \otimes \ket{\downarrow}\bra{\downarrow}+I \otimes \ket{\uparrow}\bra{\uparrow})+ \sqrt{1-p} (  E_-\otimes \ket{\downarrow}\bra{\uparrow} + E_+\otimes\ket{\uparrow}\bra{\downarrow})$
a \textit{partial} controlled-shift operation, which differs from the standard control-shift gate by also allowing the walker state to not change with some probability.
Here $\{\ket{\uparrow},\ket{\downarrow}\}$ is the computational basis for the coin space, $I$ is the identity operator on the walker space, 
$E_\pm|j\rangle \equiv |j\pm1\rangle$, with $\{\ket{j}\}$, $j=-N,...,N$ the position states of the walker, living in a $(2N+1)$-dimensional Hilbert space, and $E_-\ket{-N}=E_+\ket{N}=0$.
Finally, $\ket\psi$ is the input polarization state we seek to characterize.

After the QW evolution, the polarization is projected on some state $\ket*{\psi_{\rm pol}}$, and the OAM is measured in its computational basis.
To connect this with the general formalism introduced above, denote with $U$ the unitary corresponding to the quantum walk dynamics, $\ket*{0_{\rm OAM}}$ the initial reference OAM state. The map describing the reservoir is $\Phi(\rho)=A\rho A^\dagger$, where $A\equiv (\langle\psi_{\rm pol}|\otimes I_{\rm OAM})U(I_{\rm pol}\otimes |0_{\rm OAM}\rangle)$.
The final measurement on the OAM is then a standard projective measurement in the computational basis $\mu_b=\ketbra{b}$, with a number of outcomes that depends on the number of QW steps.

\begin{figure*}[t!]
    \centering
    \includegraphics[width=0.95\textwidth]{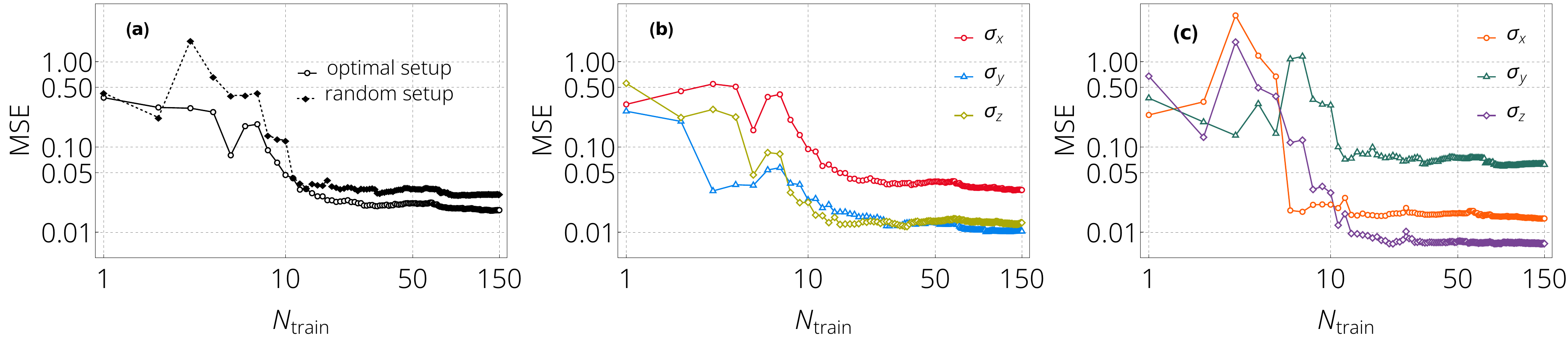}
    \caption{
    \textbf{Experimental results}.
    Estimation MSE obtained by training and testing the QELM with experimental data. The target is estimating the expectation values of the Pauli matrices, $\sigma_x,\sigma_y,\sigma_z$ on the input polarization state.
    We study the MSE as a function of the number of training states $N_{\rm train}$, at fixed statistics $N$.
    To test the protocol, we generated $300$ random input states, and tested the estimation when the first $1\le N_{\rm train} \le 150$ are used to train the QELM. The set of $300$ states remains unchanged throughout all experiments.
    The last $150$ of these $300$ states are always used for testing, to compute the MSE.
    All the points in the saturated regions of these figures  decrease as $1/N$ when increasing the statistics with which each training and test state is measured.
    \textbf{(a)} Average of the MSE estimated for all three target observables: $\{\sigma_x,\sigma_y,\sigma_z\}$.
    We show the results for both optimized and random setups.
    \textbf{(b)} MSE for each individual target observable for the optimized setup.
    \textbf{(c)} MSE for each individual target observable for the random setup.
    The reported results are obtained with average experimental statistics of $\sim 3000$ counts.
    }
    \label{fig:exp_data}
\end{figure*}

\parTitle{Optical setup}
In the experimental setup, reported in~\cref{fig:exp_setup}(\textbf{b}), a set of optical elements composed of a polarizing beam-splitter, a half-wave plate [HWP$(\zeta_1$)] and a quarter-wave plate [QWP$(\theta_1)$] produces an input polarization state parametrized as:
\begin{equation}
\begin{aligned}
    \ket\psi= \frac{1}{\sqrt{2}}
    [
    e^{i \theta_1 }(\cos{(2 \zeta_1-\theta_1)}-\sin{(2 \zeta_1 - \theta_1)})\ket L\\
    + e^{-i \theta_1 } (\cos{(2 \zeta_1-\theta_1)}+\sin{(2 \zeta_1 - \theta_1)})\ket R
    ],
\end{aligned}\label{Eq:input_state}
\end{equation}
where $\ket L$ and $\ket R$ stand for left- and right-circular polarization, $\theta_1$ and $\zeta_1$ are the rotationx angles of the waveplates optical axis.
The input state then evolves through a series of half-wave plates [HWP$(\zeta)$], quarter-wave plates [QWP$(\theta)$)], and an inhomogeneous
birefringent device, known as q-plate [QP$(\alpha,\delta)$], which couples polarization and OAM conditionally on the parameters $\delta$, the tunable phase retardance that allows the optimal tuning of the device when $\delta=\pi$, 
and $\alpha$ which is a characteristic angle associated to the initial orientation of the optical axis with respect to the horizontal direction.
QPs have been used as a building block in a significant number of demonstrations of controlled quantum dynamics~\cite{marrucci2006optical,marrucci2011spin}, and are in particular often used as controlled-shift gate to implement QW dynamics~\cite{cardano2015quantum, cardano2016statistical,giordani2019experimental,giordani2021entanglement,Gratsea_2020,suprano2021dynamical,wang2021learning}.
The coin operation is implemented via a sequence of waveplates as $C(\zeta,\theta,\phi)=\on{QWP}(\zeta)\on{HWP}(\theta)\on{QWP}(\phi)$, with $\zeta,\theta,\phi$ tunable angles.
Each q-plate implements a controlled-shift operation $S(\alpha,\delta)$ with characteristic parameters $\alpha,$ $\delta$.
More explicitly, these operations take the form
\begin{align}
    C(\zeta,\theta,\phi) =
    \begin{pmatrix}
         e^{-i(\zeta{-}\phi)}\cos\eta & e^{i(\zeta{+}\phi)}\sin\eta\\
        -e^{-i(\zeta{+}\phi)}\sin\eta & 
        e^{i(\zeta{-}\phi)}\cos\eta
    \end{pmatrix},   \\
    \begin{aligned}
        S(\alpha,\delta)=
        \sum_{n=-N+1}^{N-1}\cos\frac{\delta}{2}\left(\ket{L,n}\bra{L,n}+\ket{R,n}\bra{R,n}\right)\\
        +i \sin\frac{\delta}{2}(e^{2i\alpha}\ket{L,n}\bra{R,n{+}1}+e^{-2i\alpha}\ket{R,n}\bra{L, n{-}1}),
    \end{aligned} 
\end{align}
with $\eta=\zeta{+}\phi{-}2\theta$.\\

Here $\ket{L,n}$ ($\ket{R,n}$) denote left- (right-) circular polarization, and OAM with azimuthal quantum number $n$.
The overall evolution $U$ implemented by our apparatus is obtained by combining two controlled-shift and one coin operation:
\begin{equation}
    U= S(\alpha_2,\pi)C(\zeta,\theta,\phi)S(\alpha_1, {\pi}/{2}),
\end{equation}
where $\alpha_1,\alpha_2$ are fixed by the fabrication process, and in our case equal $105 ^\circ$ and $336 ^\circ$, respectively.
Another coin operation is used at the beginning to prepare the input state (see~\cref{Eq:input_state}), and is thus not considered as part of the reservoir dynamics.
Fixing the parameters $\delta$ of the q-plates to $\pi$ and $\pi/2$, respectively, allows us to enlarge the space of reachable output OAM states without adding QW steps, thanks to the stationary component of the dynamics.
After evolution through $U$, a combination of waveplates and a polarizing beamsplitter are used to project the polarization, while a Spatial Light Modulator (SLM) and a single-mode fiber are employed to measure the final OAM states, obtaining the occupation probabilities for the basis states $\ket{n}$, $n=-2,-1,0,1,2$.
The single-photon counts are then collected and fed to the computer, where post-processing and training of the QELM take place, and the target expectation values are estimated.

\parTitle{Results}
We considered two different configurations for the QELM. In the first, we exploited the knowledge of the QW dynamics to extract optimal values for the angles of the coin $\{\zeta, \theta, \phi\}$ and for the projection of the hidden layer which result in an almost uniform cover of the OAM space (see Supplemental Material). In the second one instead, we made a random choice of the waveplate angles, focusing on training the accessible output layer to optimize the performance of the characterization protocol.
The chosen figure of merit for the quantification of performances is the mean square error (MSE) between the expectation values of the Pauli operators. The experimental results are reported in~\cref{fig:exp_data} for both implementations. We show the performance of QELM, trained using experimental data, in retrieving the features of the polarization state. In particular, we collected $300$ experimental states and split them into a training set and a test set, each one composed of $150$ elements. The MSE of the expectation values over the test set is studied against the number $N_{train}$ of states used in the training set. A large enough training set clearly results in a decrease of the MSE, and thus in significantly enhanced reconstruction accuracies for all considered target observables.
The amount of statistics collected for each state also crucially affects the reconstruction accuracies~\cite{innocenti2022potential}. We analyze this aspect explicitly in the Supplemental Material.

Let us remark here other two significant aspects that transpire from our experimental results.
Firstly, our reconstruction protocol is highly resource-efficient
indeed roughly $20$ states are already sufficient to train the QELM, as seen in~\cref{fig:exp_data}.
Secondly, as shown in~\cref{fig:exp_data}-(\textbf{a}), the optimal configuration results in MSEs only marginally better than those obtained with the random configuration, highlighting that a full characterization and fine-tuning of the experimental setup is not essential to achieve accurate reconstruction accuracies (see details in the Supplemental Material).
 
Finally, to compare the quality of the results obtained via the QELM with non-machine-learning-based alternative approaches, we consider the reconstruction MSEs that would have been obtained with the same experimental apparatus via the generalized shadow tomography reconstruction scheme, which has been shown to be optimal for reconstruction under relatively mild assumptions~\cite{innocenti2023shadow}.
As discussed in detail in the Supplemental Materials, we find that the QELM provides performances between $5$ and $10$ times better than the alternative methods for the considered target observables, in the case of the optimal experimental setup.
A main underlying reason for this disparity is that the non-QELM-based methods rely on accurate modeling of the experimental apparatus, which can be quite costly to achieve in practice, whereas QELM can easily adapt to experimental fluctuations without significantly impacting the reconstruction accuracies.

\parTitle{Conclusions}
We have experimentally demonstrated a robust and resource-efficient QELM-based property-reconstruction protocol. Our implementation, which leverages the controlled QW dynamics in a photonic platform intertwining the polarization and OAM degrees of freedom of a photon, demonstrates the excellent performance of property reconstruction without the need for the accurate and careful characterization of the platform. Only training sets with moderate size are needed to achieve low values of the MSE of the reconstruction, while the effects of finite statistics of the dataset can be fully accounted for. Our experimental QELM-based reconstruction demonstrates the viability of photonic platforms for non-standard approaches to quantum property retrieval, with the expectation of significantly reducing the burden -- in terms of resources -- of resource-characterization in a computational register. 

\section*{Acknowledgments}
LI acknowledges support from MUR and AWS under project 
PON Ricerca e Innovazione 2014-2020, ``Calcolo quantistico in dispositivi quantistici rumorosi nel regime di scala intermedia" (NISQ - Noisy, Intermediate-Scale Quantum).
IP is grateful to the MSCA COFUND project CITI-GENS (Grant nr. 945231).
MP acknowledges the support by the European Union's Horizon 2020 FET-Open project  TEQ (Grant Agreement No.\,766900), the Horizon Europe EIC Pathfinder project  QuCoM (Grant Agreement No.\,101046973), the Leverhulme Trust Research Project Grant UltraQuTe (grant RPG-2018-266), the Royal Society Wolfson Fellowship (RSWF/R3/183013), the UK EPSRC (EP/T028424/1), and the Department for the Economy Northern Ireland under the US-Ireland R\&D Partnership Programme (USI 175 and USI 194).
AF, SL, and GMP acknowledge support by MUR under PRIN Project No. 2022FEXLYB
Quantum Reservoir Computing (QuReCo).
FS acknowledges support from the ERC Advanced Grant QU-BOSS (QUantum advantage via nonlinear BOSon Sampling, grant agreement no. 884676) and from PNRR MUR project PE0000023-NQSTI (Spoke 4).

\bibliography{bibliography}

\clearpage
\onecolumngrid
\appendix
\renewcommand{\thefigure}{S\arabic{figure}}
\renewcommand{\thetable}{S\arabic{table}}
\renewcommand{\theequation}{S\arabic{equation}}
\setcounter{figure}{0}
\setcounter{table}{0}
\setcounter{equation}{0}

\section*{Supplemental Material}

\input{SI}

\end{document}

%% file: SI.tex
\section{Theoretical background}

\subsection{QELM-based state estimation}

\parTitle{General modelization of a QELM}
Let us briefly review here the general framework to train QELMs for quantum state estimation introduced in~\cite{innocenti2022potential}.
Any measurement apparatus can be modeled via some ``effective POVM'' $\tilde{\bs\mu}\equiv(\tilde\mu_b)_{b\in\Sigma}$, where $\Sigma$ is the set of possible measurement outcomes.
If the experimental apparatus is composed of some quantum channel $\Phi$ followed by a measurement $\bs\mu\equiv(\mu_b)_{b\in\Sigma}$, the effective POVM has elements $\tilde\mu_b=\Phi^\dagger(\mu_b)$, with $\Phi^\dagger$ the adjoint of $\Phi$.
The output probabilities then have the form
\begin{equation}
    p_b(\rho) = \Trace[\mu_b\Phi(\rho)]
    = \Trace[\tilde\mu_b \rho].
\end{equation}
The input state can be recovered from the output probabilities if and only if $\tilde{\bs\mu}$ is \textit{informationally complete}, meaning that the span of the operators $\tilde\mu_b$ is the full space of Hermitian linear operators.
Performing supervised learning of a quantum extreme learning machine (QELM) involves using a training dataset of states and associated expectation values to compute an operator $W$ that can be later used to extract the target expectation values from previously unseen output probabilities.
More precisely, the training dataset has the form
$\{(\rho_k,o_k)\}_{k=1}^{N_{\rm tr}}$, with $\rho_k$ input states and $o_k=\Trace(\calO\rho_k)$ the associated expectation value of the target observable $\calO$, and $N_{\rm tr}$ the number of training states. More generally, the target can be a set of expectation values, in which case we can write the training dataset as $\{(\rho_k,\bs o_k)\}_{k=1}^{N_{\rm tr}}$, with $\bs o_k\in\mathbb{R}^\ell$ a vector with elements $(\bs o_k)_j=\Trace(\calO_j \rho_k)$, and with $(\calO_j)_{j=1}^\ell$ the target observables we mean to learn how to compute.

\parTitle{Training the QELM}
To train the QELM, each $\rho_k$ is measured $N$ times, the corresponding measurement statistics are collected into the vector of frequencies $\mathbf f_k$, and we solve for $W$ the linear system
\begin{equation}
    (W P)_{jk} = \Tr(\calO_j\rho_k),
    \qquad j=1,...,\ell,
    \quad k=1,...,N_{\rm tr}.
\end{equation}
where $P$ is the matrix whose $k$-th column is $\mathbf f_k$. Note that $P$ depends on the statistics $N$ used to estimate the outcome probabilities for each state. In the limit $N\to\infty$, its elements thus tend to the true output probabilities: $P_{bk}\to \Tr(\tilde\mu_b\rho_k)$ 
In the scenario we consider here, this linear system can be solved using the pseudoinverse, and the solution reads
\begin{equation}
    W_{jb} = 
    \sum_{k=1}^{N_{\rm tr}}
    \Tr(\calO_j \rho_k)
    (P^+)_{kb},
    \qquad P^+\equiv P^T (PP^T)^{-1},
\end{equation}
where $P^+$ is the pseudoinverse of $P$, which can always be written in the above form, as long as $P$ is surjective --- which is always the case for sufficiently many training states.

\parTitle{Application and testing of trained QELMs}
Once the training dataset has been consumed to produce $W$, this can be used to extract expectation values of any new measured input state.
In practice, this entails performing a number of measurements for a new test state $\rho$, collecting the resulting statistics into a vector $\mathbf f\equiv (f_b)_{b\in\Sigma}$ --- which in the limit of infinite statistics will equal the true probability vector $\mathbf p(\rho)\equiv \Tr[\tilde{\bs\mu}\rho]$ --- and applying the linear operator $W$ to $\mathbf f$.
We thus obtain the estimator $W\mathbf f$ for the target expectation values $\Tr(\calO_j \rho)$.
To assess the accuracy of the resulting estimate, we compute the mean squared errors (MSEs) for the target observable $\calO_j$ as
\begin{equation}
    \on{MSE}(\rho,\calO_j) \equiv \left(
    \sum_{b\in \Sigma} W_{jb} f_b - \Tr(\calO_j \rho)
    \right)^2.
\end{equation}
More precisely, the MSE is generally computed by averaging over many input states.
Because this MSE it is evaluated on the estimated frequency vectors $\mathbf f$, it is a stochastic quantity, that will have some amount of variations for finite measurement statistics.
To assess the quality of our estimation apparatus, we compute the average of $\on{MSE}(\rho,\calO_j)$ over the set of test states, thus using the overall accuracy quantifier
\begin{equation}
    \on{MSE}(\calO_j) \equiv \frac{1}{N_{\rm test}} \sum_{k=1}^{N_{\rm test}} \on{MSE}(\rho_k^{\rm test},\calO_j),
\end{equation}
denoting with $\rho_k^{\rm test}$ the set of states used to validate the performance of the method.

\parTitle{Specialization of the general notation to our photonic apparatus}
In the experimental apparatus that we consider, $\bs\mu$ is a projective measurement in the computational basis of the orbital angular momentum space, and $\Phi$ is the map describing the evolution through the photonic apparatus.
The input states $\rho$ are encoded in the polarization degree of freedom of photons, while the output states are encoded in their orbital angular momentum. Before the final measurement, the polarization is projected. This means that the overall evolution, when noise and imperfections can be neglected, has the form
\begin{equation}\label{eq:map_expression}
    \Phi(\rho) = \Tr_{\rm pol}[ (\ketbra{\psi}\otimes I_{\rm OAM}) V\rho V^\dagger], 
\end{equation}
where $V$ is the isometry that describes the overall unitary evolution of input polarization states into output polarization and orbital angular momentum states, $\ket\psi$ is the polarization state onto which we project, and $\Tr_{\rm pol}$ denotes the partial trace with respect to the polarization degree of freedom.
The final projective measurement can be written simply as $\mu_b=\ketbra{b}{b}$.

\parTitle{Nonunitarity of the evolution}
It is worth noting that, because of the polarization projection performed before the final measurement, $\Phi$ is technically not a quantum channel, and the corresponding effective POVM $\tilde{\bs\mu}$ is not a POVM.
We can see it by observing that
\begin{equation}
    \Tr[\Phi(\rho)] =
    \Tr[(\ketbra{\psi}\otimes I_{\rm OAM}) V\rho V^\dagger]
    = \langle\psi| \Tr_{\rm OAM}[V\rho V^\dagger] |\psi\rangle \le \Tr(\rho),
\end{equation}
where the inequality is generally \textit{not} saturated, unless $\Tr_{\rm OAM}[V\rho V^\dagger]=\ketbra{\psi}$, which in turn is only possible if $V\rho V^\dagger$ has no entanglement whatsoever between polarization and orbital angular momentum.
In other words, as long as the evolution entangles polarization and orbital angular momentum --- which is a necessary requirement to enable reconstruction --- $\Phi$ does not preserve the trace, and is thus not a quantum channel.
In terms of the effective POVM, this condition is reflected on the normalization condition being weakened to $\sum_b\tilde\mu_b\le I$, rather than being an identity.
Nonetheless, $\Phi$ not being trace-preserving does not affect the QELM reconstruction procedure previously discussed, and its only practical effect is an overall reduction of observed statistics.
We can therefore safely neglect this detail and proceed as if $\Phi$ was a proper quantum channel.

\subsection{Measurement frames and optimal measurement}
\label{sec:frames_and_optimal_measurement}

\parTitle{QELMs- vs frame-based reconstruction}
One of the big advantages of the QELM-based approach to state estimation is that it does not require any prior knowledge of the effective POVM $\tilde{\bs\mu}$ implemented by the experimental apparatus.
This means that even if our theoretical model for the apparatus is not entirely accurate, this will not hinder the estimation accuracies.
On the other hand, if we \textit{do} have knowledge of $\tilde{\bs\mu}$, that is, we have an accurate model for the experimental apparatus, then we can exploit it to obtain precise estimates of the expected estimation accuracies in any given scenario.
We will here briefly review the formalism introduced in~\cite{innocenti2023shadow} to analyze estimation errors for arbitrary target observables and POVMs.
Given any POVM $\tilde{\bs\mu}$ and target observable $\calO$, we can build an estimator for $\Tr(\calO\rho)$ as the function $b\mapsto \Tr(\calO\tilde\mu_b^\star)$, where $\tilde\mu_b^\star$ are the elements of the \textit{dual measurement frame}, defined as
\begin{equation}
    \tilde\mu_b^\star = S^{-1}(\tilde\mu_b),
    \qquad
    S(X)\equiv \sum_{b\in\Sigma} \Tr(\tilde\mu_b X) \tilde\mu_b.
\end{equation}
This estimator is \textit{unbiased}, meaning it gives the correct value of $\Tr(\calO\rho)$ on average.
There are other possible choices of unbiased estimators. A notable one being the so-called \textit{canonical estimator}, which is the function $b\mapsto \Tr(\calO\tilde\mu_b^{\rm can})$ defined as
\begin{equation}\label{eq:canonical_frame_superop}
    \tilde\mu_b^{\rm can} = \frac{F^{-1}(\tilde\mu_b)}{\Tr(\tilde\mu_b)/d},
    \qquad
    F(X)\equiv \sum_{b\in\Sigma} \frac{\Tr(\tilde\mu_b X)\tilde\mu_b}{\Tr(\tilde\mu_b)/d},
\end{equation}
with $d$ the dimension of the underlying state space.
As discussed in detail in~\cite{innocenti2023shadow}, this estimator is ``optimal'' in the sense of being the unbiased estimator for $\Tr(\calO\rho)$ with the lowest possible variance when no prior knowledge on $\rho$ is assumed.
Estimation strategies based on such estimators can be shown to be optimal, for the information accessible with the given effective POVM, but need to assume prior knowledge of $\bs{\tilde\mu}$.

\parTitle{Estimation error}
The estimation accuracy associated to any given estimator can be quantified via its variance. In particular, using the canonical estimator for an observable $\calO$, the variance is given by the expression:
\begin{equation}\label{eq:avg_variance}
    \overline{\on{Var}[\hat o]} =
    \langle \calO,F^{-1}(\calO)\rangle
    - \beta,
    \qquad
    \beta\equiv \frac{\Tr(\calO)^2}{d^2}
    + \frac{dP-1}{d^2-1} V,
    \qquad
    V \equiv \frac{\Tr(\calO^2)}{d}
    - \frac{\Tr(\calO)^2}{d^2},
\end{equation}
where the constant $\beta$, crucially, does not depend on the choice of measurement.
More specifically, $\overline{\on{Var}[\hat o]}$ is the variance of $\hat o$, averaged over all possible input states.
On the other hand, $\langle\calO,F^{-1}(\calO)\rangle$ depends on $\bs\mu$ through $F$.


\parTitle{Optimal measurement strategies}
We can use the framework thus described to find the quantum walk evolution that results in a map $\Phi$ corresponding to the effective POVM $\bstildemu$ with the smallest possible averaged variance.
Following~\cref{eq:avg_variance}, this amounts to minimizing $\langle\calO,F^{-1}(\calO)\rangle$ with respect to the possible measurement choices. We can further simplify this task by removing the dependence on the observable $\calO$ and minimizing instead $\Tr(F^{-1})$.
As shown in~\cite{scott2006tight,zhu2011quantum,innocenti2023shadow,haah2016sample}, this amounts to finding the effective measurement which provides the minimal averaged state estimation error in the Hilbert-Schmidt distance.
To perform this minimization, we first obtain a parametrization of the effective POVM in terms of the evolution parameters. From~\cref{eq:map_expression} we see that $\bstildemu$ has the expression
\begin{equation}\label{eq:effective_POVM}
    \tilde\mu_b = V^\dagger (\ketbra{\psi}\otimes \mu_b)V.
\end{equation}
As previously mentioned, this is not strictly speaking a POVM because it is not normalized, but our formalism works regardless of this detail.
The parameterization of $\bstildemu$ then passes through that of $V$ and $\ket\psi$.
In other words, because in our apparatus $V$ is implemented as a sequence of properly tuned waveplates and quplates, each one of which is characterized by some adjustable parameters, we can explore the space of possible isometries $V$ achievable with our apparatus.


\section{Optimal measurement apparatus}
\label{sec:optimal_measurement}

\parTitle{General strategy}
We discussed in~\cref{sec:frames_and_optimal_measurement} how the formalism of measurement frames can be leveraged to compute the expected estimation accuracies in any given measurement scenario.
Furthermore,~\cref{eq:effective_POVM} gives us the explicit form of the effective POVM corresponding to our experimental platform.
Finding the experimental parameters that provide an optimal estimation strategy then amounts to finding the isometry $V$ and projection $\ket\psi$ that minimize $\Tr(F^{-1})$, with $F$ defined as in~\cref{eq:canonical_frame_superop}.
We will provide here further details on the explicit form of these operators, and the optimization process.

\parTitle{Modelization of optical elements}
As mentioned in the main text, in the experiment we use wave-plates, characterized by the parameters $\zeta,\theta,\phi\in\mathbb{R}$, and q-plates $S(\alpha,\delta)=\text{QP}(\alpha,\delta)$, with $\alpha,\delta\in\mathbb{R}$.
The waveplates are used to implement coin operations on the polarization, $C(\zeta,\theta,\phi)=\on{QWP}(\zeta)\on{HWP}(\theta)\on{QWP}(\phi)$, with $\zeta,\theta,\phi$ the angles defining the rotation of the polarization state,
and each q-plate implements a controlled-shift operation.
The unitaries corresponding to these operations take the form
\begin{equation}
\begin{gathered}
C(\zeta,\theta,\phi) =
    \begin{pmatrix}
         e^{-i(\zeta{-}\phi)}\cos\eta & e^{i(\zeta{+}\phi)}\sin\eta\\
        -e^{-i(\zeta{+}\phi)}\sin\eta & 
        e^{i(\zeta{-}\phi)}\cos\eta
    \end{pmatrix},
    \\
    S(\alpha,\delta) =\sum_{n=-N+1}^{N-1}\cos\frac{\delta}{2}\left(\ket{L,n}\bra{L,n}+\ket{R,n}\bra{R,n}\right)
    +i \sin\frac{\delta}{2}(e^{2i\alpha}\ket{L,n}\bra{R,n{+}1}+e^{-2i\alpha}\ket{R,n}\bra{L, n{-}1}), 
\end{gathered}
\end{equation}

with $\eta=\zeta{+}\phi{-}2\theta$. Here $\ket{L,n}$ ($\ket{R,n}$) denote left-circular (right-circular) polarization, and OAM with azimuthal quantum number $n$.
More specifically, our apparatus corresponds to the unitary evolution
$
    U= S(\alpha_2,\pi)C(\zeta,\theta,\phi)S(\alpha_1, {\pi}/{2})
$, 
with parameters $\alpha_1,\alpha_2,\zeta,\theta,\phi\in\mathbb{R}$. This unitary is then connected to the isometry $V$ via $V=U(I_{\rm pol}\otimes \ket{0_{\rm OAM}})$, which thus maps two-dimensional input polarization states into five-dimensional OAM ones.
We can further simplify our modelization of the apparatus by integrating the final projection into the isometry.
Defining the linear operator $\tilde V\equiv ( \bra{\psi}\otimes I_{\rm OAM})V$, 
we can concisely write the output state as $\tilde V\rho \tilde V^\dagger$, with $\tilde V$ the operator characterized by the tunable parameters. 
Explicitly, $\tilde V$ has the expression
\begin{equation}
\tilde V = \frac{1}{\sqrt{2}}
\begin{pmatrix}
 0 & e^{2 i \alpha_1+2 i \alpha_2-i \zeta-i \phi}\sin (\eta )\cos(\theta_p) \\
 -i e^{2 i \alpha_2-i \zeta-i \phi} \sin(\eta)\cos(\theta_p) & ie^{2 i \alpha_2+i \zeta-i \phi} \cos (\eta)  \cos(\theta_p) \\
 -e^{-2 i \alpha_1+2 i \alpha_2+i \zeta-i \phi}\cos(\eta)\cos(\theta_p) & -e^{2 i \alpha_1-2 i \alpha_2-i \zeta+i \phi_p+i \phi}\cos (\eta) \sin(\theta_p)  \\
 i e^{-2 i \alpha_2-i \zeta+i \phi_p+i \phi} \cos (\eta) \sin(\theta_p) & i e^{-2 i \alpha_2+i \zeta+i \phi_p+i \phi}  \sin (\eta) \sin(\theta_p)\\
 -e^{-2 i \alpha_1-2 i \alpha_2+i \zeta+i \phi_p+i \phi} \sin (\eta) \sin(\theta_p)  & 0
\end{pmatrix},
\end{equation}
where the polarization projection is characterized by the angles $\theta_p,\phi_p$ via  $\ket\psi=\cos(\theta_p)\ket0+\sin(\theta_p)e^{i\phi_p}\ket1$.

\begin{figure}[tbh]
    \centering
    \includegraphics[width=0.75\linewidth]{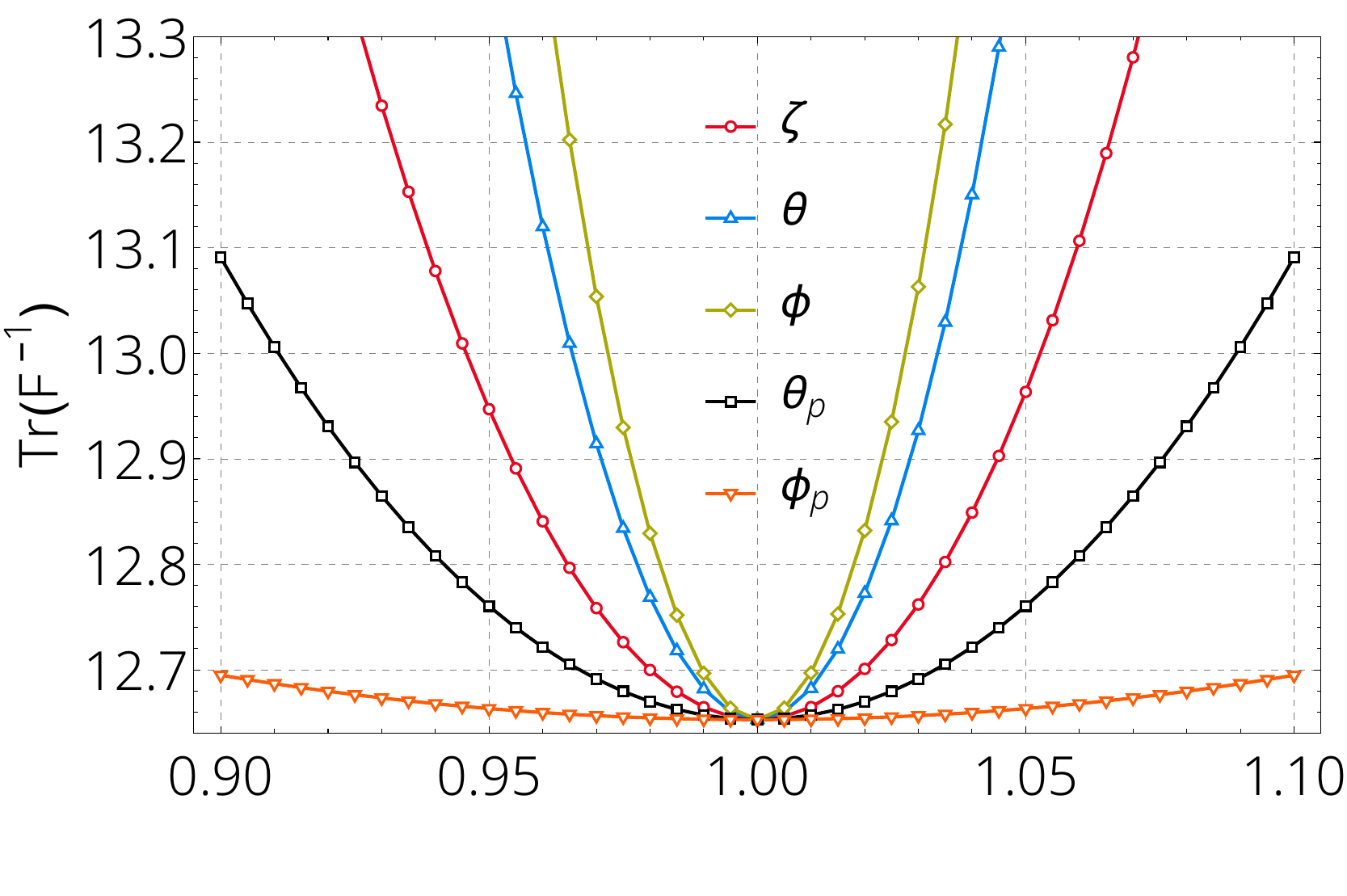}
    \caption{
    \textbf{Smoothness of reconstruction variance}
    We show the average estimation variance with respect to small variations of individual parameters characterizing the experimental apparatus.
    Each parameter is perturbed independently, leaving all the other parameters fixed at their optimal value.
    This showcases how different parameters affect the overall estimation accuracies in possibly different ways.
    }
    \label{fig:stability_trFm1_vs_parameters}
\end{figure}

\parTitle{Optimization of experimental parameters}
From $\tilde V$ we compute the five operators in $\bstildemu$, which read
\begin{eqnarray}
&\tilde\mu_1=\frac{1}{2}\begin{pmatrix}
 0 & 0 \\
 0 & S_{\eta}^2 C_{\theta_p}^2
\end{pmatrix}\qquad 
\tilde\mu_2=\frac{1}{2}\begin{pmatrix}
S_{\eta}^2 S_{\theta_p}^2 & 0 \\
0 & 0
\end{pmatrix}\qquad 
\tilde\mu_3=\frac{1}{2}\begin{pmatrix} 
 S_{\eta }^2 C_{\theta _p}^2 & -\frac{e^{-2 i \zeta}}{2} S_{2\eta } C_{\theta _p}^2 \\
-\frac{e^{2 i \zeta}}{2} S_{2\eta } C_{\theta _p}^2 & C_{\eta }^2 C_{\theta _p}^2 \\
\end{pmatrix}\qquad \\
\nonumber
& \tilde\mu_4=\frac{1}{2}\begin{pmatrix} 
 C_{\eta }^2 C_{\theta _p}^2 & \frac{e^{-2 i \nu}}{2} C_{\eta }^2 S_{2\theta _p} \\
 \frac{e^{2 i \nu}}{2} C_{\eta }^2 S_{2\theta _p} & C_{\eta }^2 S_{\theta _p}^2 \\
\end{pmatrix}
\qquad
\tilde\mu_5=\frac{1}{2}\begin{pmatrix}
 C_{\eta }^2 S_{\theta _p}^2 & \frac{e^{-2 i \zeta}}{2} S_{2\eta } S_{\theta _p}^2 \\
 \frac{e^{2 i \zeta}}{2} S_{2\eta } S_{\theta _p}^2 & S_{\eta }^2 S_{\theta _p}^2
\end{pmatrix},
\end{eqnarray}
where we introduced the auxiliary variable $\nu=4\alpha_1-4 \alpha_2-2\zeta+\phi_p+2\phi$ for notational brevity, and used the shorthand notation $C_\alpha{\equiv}\cos(\alpha)$, $S_\alpha{\equiv}\sin(\alpha)$.
From these we compute the frame superoperator $F$ via~\cref{eq:canonical_frame_superop}, obtaining:
\begin{equation}\small
F =
\frac{1}{2} \begin{pmatrix}
 2 & C_{\eta }^2 C_{\nu } S_{2 \theta _p}-C_{2 \zeta } S_{2 \eta } C_{2 \theta _p} & C_{\eta }^2 S_{\nu } S_{2 \theta _p}-S_{2 \zeta } S_{2 \eta } C_{2 \theta _p} & 0 \\
 C_{\eta }^2 C_{\nu } S_{2 \theta _p}-C_{2 \zeta } S_{2 \eta } C_{2 \theta _p} & C_{\eta }^2 C_{\nu }^2 S_{2 \theta _p}^2+C_{2 \zeta }^2 S_{2 \eta }^2 & \frac{1}{2} \left(C_{\eta }^2 S_{2 \nu } S_{2 \theta _p}^2+S_{4 \zeta } S_{2 \eta }^2\right) & \frac{1}{2} \left(C_{\eta }^2 C_{\nu } S_{4 \theta _p}+C_{2 \zeta } S_{4 \eta }\right) \\
 C_{\eta }^2 S_{\nu } S_{2 \theta _p}-S_{2 \zeta } S_{2 \eta } C_{2 \theta _p} & \frac{1}{2} \left(C_{\eta }^2 S_{2 \nu } S_{2 \theta _p}^2+S_{4 \zeta } S_{2 \eta }^2\right) & C_{\eta }^2 S_{\nu }^2 S_{2 \theta _p}^2+S_{2 \zeta }^2 S_{2 \eta }^2 & \frac{1}{2} \left(C_{\eta }^2 S_{\nu } S_{4 \theta _p}+S_{2 \zeta } S_{4 \eta }\right) \\
 0 & \frac{1}{2} \left(C_{\eta }^2 C_{\nu } S_{4 \theta _p}+C_{2 \zeta } S_{4 \eta }\right) & \frac{1}{2} \left(C_{\eta }^2 S_{\nu } S_{4 \theta _p}+S_{2 \zeta } S_{4 \eta }\right) & C_{2 \eta }^2+C_{\eta }^2 C_{2 \theta _p}^2+S_{\eta }^2 \\
\end{pmatrix}.
\end{equation}
Finally, we find the optimal values of the parameters by numerically minimizing $\Tr(F^{-1})$ with respect to $\theta_p,\phi_p,\zeta,\theta,\phi$, while 
$\alpha_1=105^\circ$, $\alpha_2=336^\circ$ are left fixed.
The values of the parameters resulting from the optimization are
\begin{equation} 
    \zeta=0.82223, 
    \quad
    \theta=1.14266, 
    \quad
    \phi=2.26421,
    \quad
    \theta_p=0.78539, 
    \quad
    \phi_p=0.75016.
\end{equation}
In~\cref{fig:stability_trFm1_vs_parameters} we show the variation of $\Tr(F^{-1})$ around the optimal value when varying each parameter individually, to assess the sensitivity of the averaged reconstruction MSE with respect to small parameter changes
.


\section{Effects of experimental imperfections}

\parTitle{Detailed description of the measurement strategy}
The quantum walk architecture we employ generates, after the polarization projection, superpositions of OAM states spanning $5$ possible basis states, labeled as $\{-2,-1,0,1,2\}$.
The population on each output OAM state $\ket m$ is then measured using an SLM to rotate $\ket m$ on the TEM$_{00}$ state, which is then coupled on a single-mode fiber (SMF), and the resulting statistics measured with an avalanche photodiode detector (APD).
We performed rounds of measurements with $300$ different input states and, to study how estimation accuracies vary with the statistics, this whole acquisition procedure was repeated $12$ times --- we will refer to these repetitions as \textit{batches} in the following.
Specifically, to measure each state in each batch, we projected, via an SLM, the output OAM mode onto the elements of the computational basis $\{-2,-1,0,1,2\}$ and for each projection we collected data for an acquisition time of $\sim 4$s.
Note that due to the polarization projection probability depending on the input state, the constant acquisition time results in different numbers of observed events for different states.
The same set of states is used in each of these $12$ batches so that the associated statistics can be combined to study the performance as a function of the acquired statistics.
Each of these $12$ batches of data is split into $150$ training and $150$ testing states, used for training and testing, respectively.
We test the reconstruction strategy to recover the expectation values of the  Pauli matrices, $\{\sigma_x,\sigma_y,\sigma_z\}$, on each of the testing states, using the MSE to quantify accuracies.
This whole procedure is repeated with two different experimental setups, one with randomly selected waveplates and projection angles, and the other using the procedure outlined in~\cref{sec:optimal_measurement} to find the angles that in the theoretical model provide optimal estimation accuracies.

\begin{figure*}[htb!]
    \centering
    \includegraphics[width=0.95\textwidth]{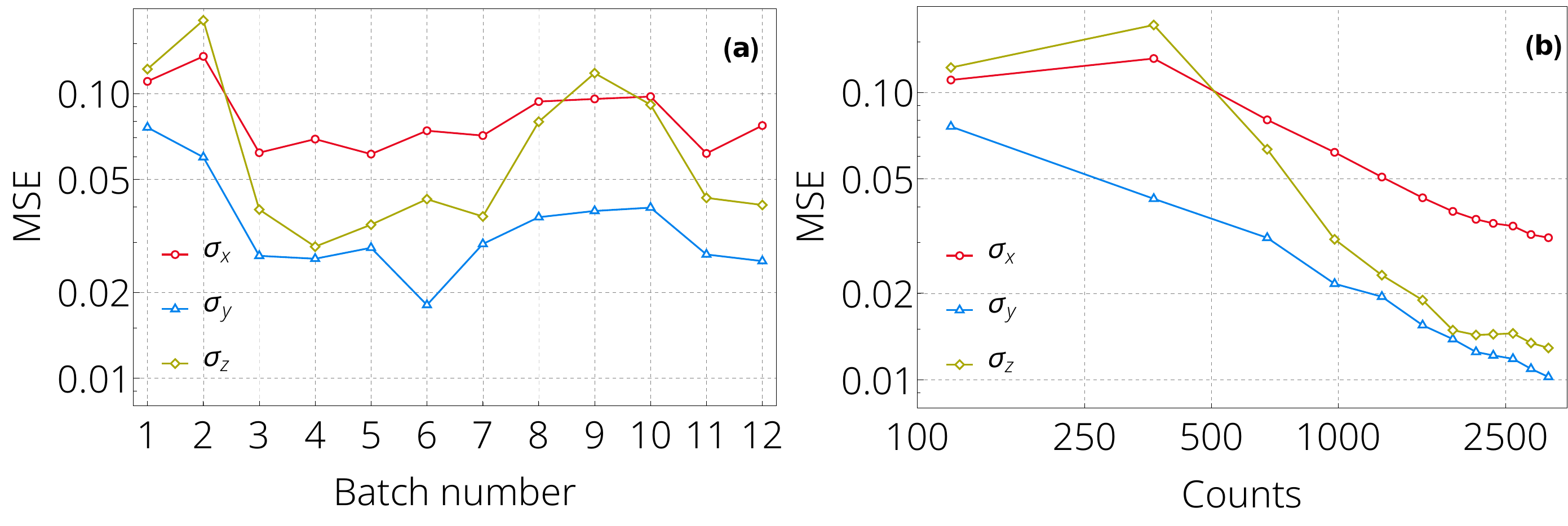}
    \caption{\textbf{MSE vs statistics}.
    Reconstruction MSE for the observables $\sigma_x,\sigma_y,\sigma_z$ averaged over the training states, measuring states with the optimal experimental configuration.
    This data corresponds to the optimal measurement setup described in~\cref{sec:optimal_measurement}.
    \textbf{(a)} MSE for each of the three target observables, for each batch of collected data. Each batch contains $300$ states, half of which are used to train the QELM, and the remaining half is used to compute the reported MSE. Each batch contains the experimental counts obtained measuring the same set of $300$ states, and they therefore only differ due to statistical fluctuations and thermal fluctuations potentially affecting the alignment of the apparatus.
    Training and testing are performed independently in each batch.
    \textbf{(b)}
    MSE for each of the three target observables, where we merge the data in the $12$ batches to study how the MSE changes with the amount of collected statistics.
    We cumulatively merge the statistical data for each of the $300$ states used in all $12$ batches, thus simulating an experiment where each of the states has been measured with longer and longer acquisition times.
    The training and testing is then performed on the resulting data.
    }
    \label{fig:mse_vs_statistics}
\end{figure*}

\begin{figure}[htb]
    \centering
    \includegraphics[width=0.8\linewidth]{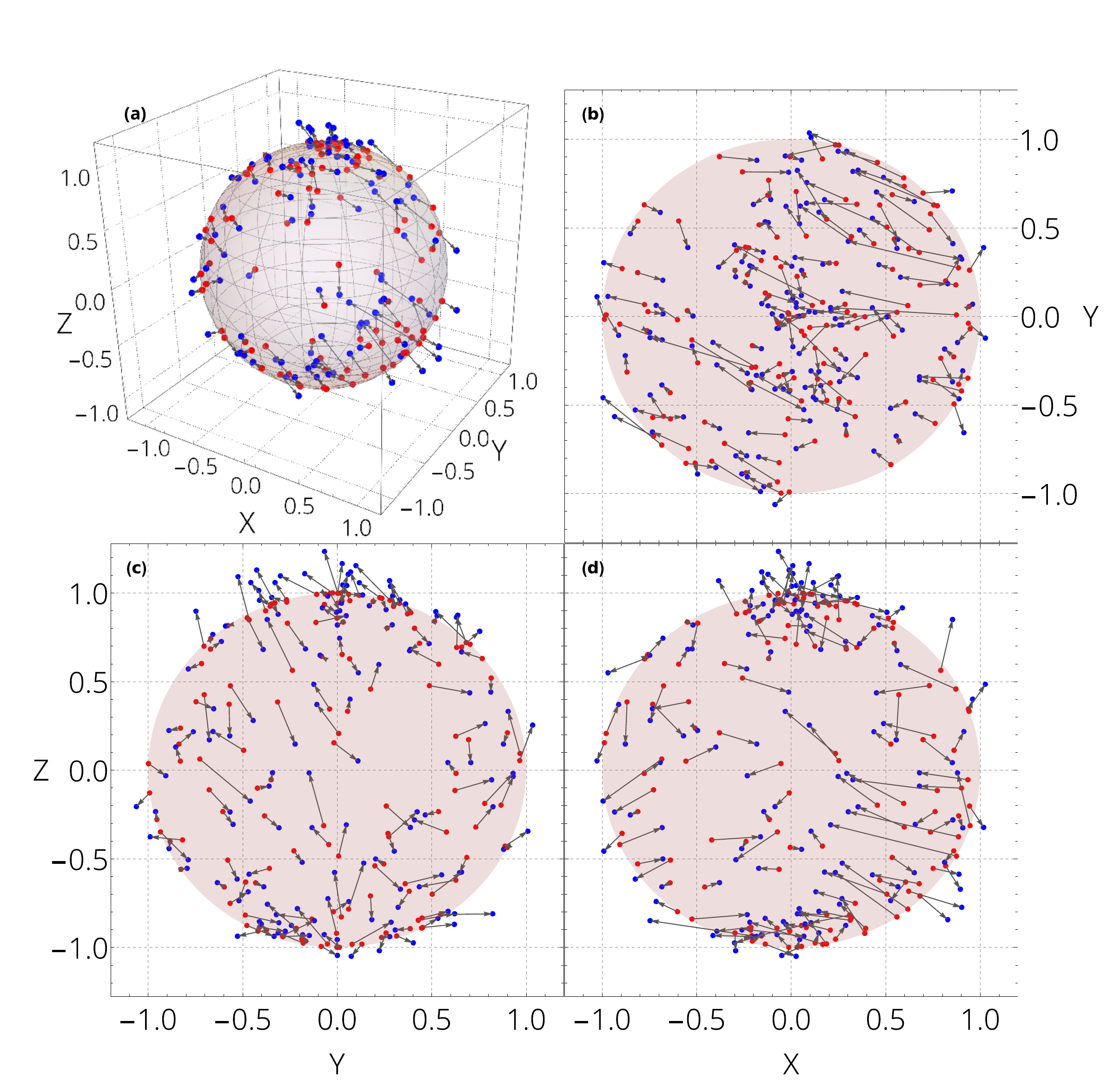}
    \caption{
    \textbf{Estimated vs true expectation values.}
    Direct comparison between true and estimated expectation values for the three Pauli observables.
    For each input state $\rho$, we show in red the point $(\Tr(\sigma_x \rho),\Tr(\sigma_y \rho),\Tr(\sigma_z \rho))\in\mathbb{R}^3$, while the connected blue point at the end of each arrow shows the reconstruction with the trained QELM.
    The data shown here corresponds to the first $100$ test states, using all training states to train the QELM, with the experimental data obtained with the optimal configuration discussed in~\cref{sec:optimal_measurement}, using all the available statistics obtained with the $12$ batches.
    }
    \label{fig:estimated_vs_reconstructed_on_sphere}
\end{figure}

\parTitle{Limitations of experimental setup}
The limitations linked to experimental imperfection on the QELM reconstruction can be categorized into two main types. Firstly, the constraint of limited statistics directly impacts the accuracy of our final estimations. The finite number of measurements available significantly influences the reconstruction process within the QELM framework. Secondly, the apparatus is susceptible to experimental fluctuations induced by both thermal and mechanical instabilities, as well as broader imperfections stemming from suboptimal calibration of optical components. These fluctuations can introduce variability in experimental conditions, thereby affecting the reconstructed states.
To assess the stability of the apparatus we analyze the MSE on each individual batch of data.
In~\cref{fig:mse_vs_statistics}-(a) we report the MSE on the test states for each individual batch, when we also perform the QELM training independently in each batch. The observed fluctuations in the MSE are attributed to natural statistical fluctuation, causing perturbations in the experimental apparatus between different batches.
In~\cref{fig:mse_vs_statistics}-(b) we instead consider the acquired data cumulatively for each state.
In this way, we can study how the estimation accuracy for each state changes with the amount of statistics used to measure it.
This shows clearly the expected downward trend of the MSE with the collected statistics.
These results showcase that even in the presence of instabilities in the experimental apparatus, which would cause reconstruction with standard techniques to not perform well due to systematic errors, the QELM-based approach can automatically account for such imperfections and still provide accurate results obviating the need for exhaustive characterization of all experimental imperfections.
To gain a more intuitive insight into the kind of reconstruction accuracies obtained via this protocol we also show in~\cref{fig:estimated_vs_reconstructed_on_sphere} the closeness of true and reconstructed states in the Bloch sphere.

\parTitle{Possible sources of MSE saturation}
It is worth noting that, although in theory the MSE will decrease indefinitely with $1/N$, this holds under the assumption that the overall dynamics can be described as some fixed quantum channel $\Phi$ or, equivalently, a fixed effective measurement $\bs{\tilde\mu}$. Fluctuations in the experimental apparatus can violate this assumption and induce a lower bound on the achievable MSE. However, as shown in Figure S2 (\textbf{b}), we find that, through the full range of collected statistics, we do not reach such a regime.
    We expect that, by further increasing $N$, the MSE will at some point stop decreasing due to fluctuations in the experimental apparatus and measurement, and saturate to a value determined by the magnitude of such fluctuations and that is difficult to estimate {\it a priori}.

\parTitle{Error propagation}
To estimate the expected estimation error we observe that the estimators obtained training the QELM are linear functionals of the output probabilities, meaning that the expectation value of a target observable $\mathcal O$ is estimated as $\hat o_N = \sum_b \hat p(b)_N \hat o_b$. Here, $\hat p(b)_N$ is the probability of observing the $b$-th outcome estimated with statistics $N$, and $\hat o_b\equiv w_b$ the elements of the linear operator obtained training the QELM.
Consequently, the expected magnitude of the estimation errors is obtained via standard statistical considerations, for example from the variance of $\hat o_N$.
More concretely, this would involve computing the sample variance as
\begin{equation}
    \sigma^2 = \frac{1}{N-1} \left[ \overline{o^2} - \overline{o}^2 \right],
    \qquad
    \overline o\equiv \frac{\sum_{k=1}^N \hat o_{b_k}}{N},
    \quad
    \overline{o^2}\equiv \frac{\sum_{k=1}^N \hat o_{b_k}^2}{N},
\end{equation}
where $\overline o$ is the sample mean of the estimated expectation value, and $\overline{o^2}$ its squared mean, and $\hat o_{b_k}$ is the value of the expectation value estimated from the $k$-th experimental observation.
This method can be effortlessly applied to obtain error bounds for the expectation value of any target observable for any input state.
We do not explicitly use this approach in the manuscript as we are interested in typical values of the MSE across many input states.
In such a scenario, it is more convenient to instead directly study the average MSE, which we can do because in, in our proof-of-principle experiment, we know the true expectation values to estimate.
If one were to instead apply our approach to estimate properties of truly unknown input states, the previously outlined method involving the sample variance would be used.

%

\section{Comparison with alternative reconstruction method}
\label{sec:comparison_standard_method}

\parTitle{Non-machine-learning based estimation as a benchmark}
To assess the quality of the proposed estimation method in our specific experimental scenario, a natural benchmark is the performances we would have obtained using a standard non-machine-learning-based approach.
Being our focus the estimation of target observables, the natural approach is arguably the framework for linear reconstruction of target observables for arbitrary POVMs~\cite{haah2016sample,zhu2011quantum,scott2006tight,bisio2009optimal}.
In particular, we can employ the formalism recently introduced by some of us~\cite{innocenti2023shadow} to devise the optimal reconstruction procedure associated with any given measurement apparatus, and analyze the resulting expected estimation errors, that we briefly introduced in~\cref{sec:frames_and_optimal_measurement}.
The gist of the technique is to define the estimator for the target observable via the canonical dual frame associated with the effective POVM $\bstildemu$, as in~\cref{eq:canonical_frame_superop}.
An issue is, however, that this way of computing the estimator relies on the knowledge of $\bstildemu$, which in practice means having a very accurate model of the measurement apparatus.
Any error in such modelization will seep into the associated estimation errors.
By contrast, QELMs do not require any such prior knowledge, trading it for the need to access a pre-characterized training dataset, which in many scenarios is easier to achieve.

\parTitle{Details of the comparison}
We benchmark the results obtained with the QELM for the optimal experimental scenario described in~\cref{sec:optimal_measurement}. This involves using~\cref{eq:effective_POVM} to compute $\bstildemu$ using the suitable choices of isometry $V$ and polarization projection $\ket\psi$, and then leveraging~\cref{eq:canonical_frame_superop}, compute the minimum-variance unbiased estimator for $\calO$ as $\Tr(\calO\tilde\mu_b^{\rm can})$.
If a previously unseen test state $\rho$ is then evolved and measured $N$ times, with the $b$-th outcome found $N_b$ times, then the estimate for $\Tr(\calO \rho)$ is given by
\begin{equation}\label{eq:observable_estimator}
    \hat o = \sum_{b\in\Sigma} \tilde\mu_b^{\rm can} \frac{N_b}{N} .
\end{equation}
As already mentioned, the optimality of this estimator hinges on the assumption that $\bstildemu$ perfectly models the actual experimental apparatus. An additional requirement is knowledge of the total number of measured states $N$. 
The latter can also be an issue in practice because, in lack of heralding of the photon sources, we do not know precisely the amount of photons entering the apparatus that were lost to noise and decoherence, and thus not observed at the output.
Note that, again, QELMs do not have this issue, as knowledge of $N$ is not required, and the final estimator is derived automatically from the training dataset.

\begin{figure}[h!]
    \centering
    \includegraphics[width=0.75\linewidth]{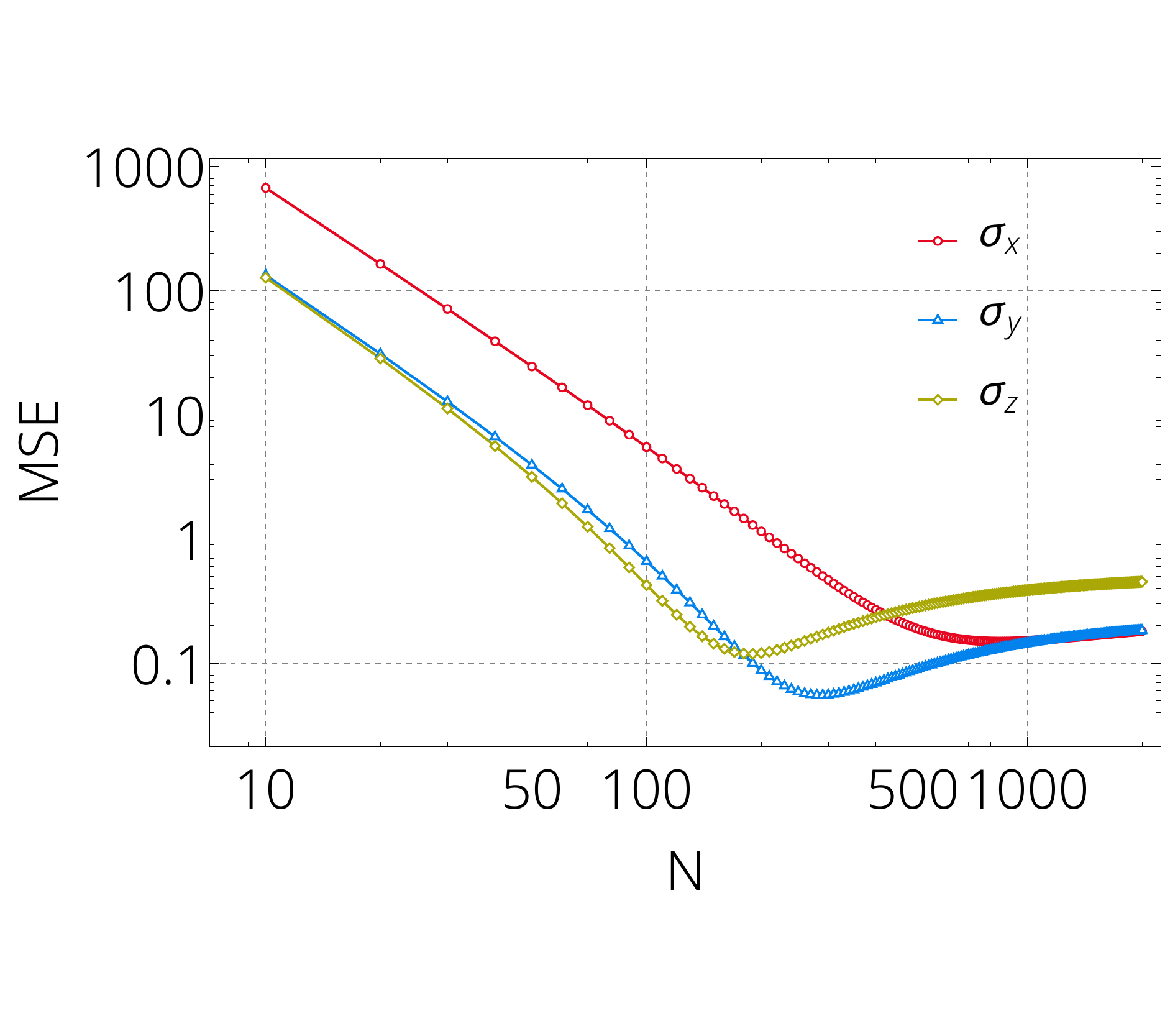}
    \caption{
    Reconstruction MSE for the Pauli observables, using the shadow tomography approach, as a function of the prior total statistics $N$ used to compute the estimator with~\cref{eq:observable_estimator}.
    The shadow tomography approach relies on knowledge of the effective POVM $\bstildemu$ and the total number of input states $N$, but the experimental apparatus we consider does not give direct access to $N$, as we cannot fully discern photon losses from a lower total number of photons injected at the input.
    The MSE for all observables remains higher than $0.05$ in all cases. In particular, for $\sigma_y$ the minimum is $0.055$ at $N=290$, for $\sigma_x$ the minimum is $0.15$ at $N=850$, and for $\sigma_z$ the minimum is $0.12$ at $N=190$.
    }
    \label{fig:MSE_vs_N_frame}
\end{figure}

\parTitle{Results}
The effective POVM elements $\tilde\mu_b$ corresponding to the optimal experimental apparatus are those given in~\cref{sec:optimal_measurement}. To overcome our lack of knowledge of $N$, we considered the minimum variance we would get among all possible values of $N$, in order to obtain a best-case-scenario performance to compare with those given by the QELM.
\Cref{fig:MSE_vs_N_frame} shows the behaviour of the MSE as a function of $N$, using the Pauli matrices as target observables.
We can then take the $N$ minimizing such MSE, and use it to build our estimator with~\cref{eq:observable_estimator}.
It is important to stress that this provides a very optimistic performance estimate, for a number of reasons:
(1) it is not true in general that the true $N$ is the one minimizing the MSE thus computed; (2) as clearly seen from the figure, the $N$ minimizing the MSE changes with the target observable, whereas the true $N$ is a unique observable-independent value; (3) in some cases the $N$ minimizing the MSE is observed to have very large values, sometimes $N\to\infty$, or anyway much larger than the statistics realistically present in the experiment. This can happen when the $\bstildemu$ used in defining $\hat o$ is not very accurate, making the trivial estimator $\hat o=0$ better than the one we would have obtained via the ``smart'' estimation procedure.
These are all issues arising from an imprecise modeling of the apparatus.
We nonetheless employ these strategies to obtain a very optimistic assessment of the accuracies obtained with this method, as even in this regime the performances are significantly worse than those obtained via the QELM.
In particular, the MSE thus obtained remains \textit{higher} than $0.05$ in all cases, whereas with the QELM, as can be seen from Figure 2 in the main text, the reconstructed MSEs are always \textit{lower} than $0.05$. In particular, with the trained QELM for $\sigma_y$ and $\sigma_z$ the MSE is lower than $0.015$, whereas with the shadows approach the MSE is higher than $0.1$. We similarly observe a roughly fivefold performance improvement with the QELM in the case of $\sigma_x$.
These differences would significantly increase when using a single value of $N$ for all observables, which would also be the approach one would have to take to make comparisons of the tomography reconstruction performances in terms of fidelity or trace distance.
\newpage

